\documentclass[fleqn,usenatbib]{mnras}

\usepackage{newtxtext,newtxmath}

\usepackage[T1]{fontenc}
\usepackage{ae,aecompl}

\usepackage{graphicx}	
\usepackage{amsmath}	
\usepackage{amssymb}	
\usepackage{color}

\newcommand{\green}[1]{{\color{\green} #1}}

\title[Intrinsic polarisation of historical Sgr A* flares]{Impact of intrinsic polarisation of Sgr~A* historical flares \\ on (polarisation) properties of their X-ray echoes}
\author[Khabibullin, Churazov \& Sunyaev]{
Ildar Khabibullin$^{1,2}$,
Eugene Churazov$^{1,2}$,
and Rashid Sunyaev$^{1,2}$
\\\\
$^{1}$ MPI f\"ur Astrophysik, Karl-Schwarzschild str. 1, Garching D-85741, Germany\\
$^{2}$ Space Research Institute, Profsoyuznaya str. 84/32, Moscow, 117997, Russia\\
}

\date{Accepted XXX. Received YYY; in original form ZZZ}

\pubyear{2020}

\begin{document}
\label{firstpage}
\pagerange{\pageref{firstpage}--\pageref{lastpage}}
\maketitle

\begin{abstract}
Reflection of X-ray emission on molecular clouds in the inner $\sim$ 100 pc of our Galaxy reveals that, despite being extremely quiet at the moment, our supermassive black hole Sgr\,A* should have experienced bright flares of X-ray emission in the recent past. Thanks to the improving characterisation of the reflection signal, we are able to infer parameters of the most recent flare(s) (age, duration and luminosity) and relative line-of-sight disposition of the brightest individual molecular complexes. We show that combining these data with measurements of polarisation in the reflected X-ray continuum will not only justify Sgr\,A* as the primary source but also allow deriving intrinsic polarisation properties of the flare emission. This will help {to identify} radiation mechanisms and underlying astrophysical phenomena behind them. For the currently brightest reflecting molecular complex, Sgr A, the required level of sensitivity might be already accessible with upcoming X-ray polarimeters.
\end{abstract}
\begin{keywords}
X-rays: general --  ISM: clouds --  galaxies:  nuclei  -- Galaxy: centre -- X-rays: individual: Sgr A* -- radiative transfer
\end{keywords}

\section{Introduction}
\label{s:introduction}

~~~~~~Due to very low accretion rate (estimated at level $10^{-8}$ of the critical value), the observed emission from our Galaxy's central supermassive black hole (SMBH) Sgr~A* is both very dim and highly variable {\citep[][and references therein]{2001Natur.413...45B,2003Natur.425..934G,2019ApJ...871..161B}}. In particular, it features remarkable high-amplitude flares (a factor of $\sim$ few hundred relative to the quiescent level) lasting for a few hours {\citep[e.g.][]{2019ApJ...886...96H}}. Extensive IR and X-ray monitoring of these flares allowed one to probe their spectral and temporal characteristics {\citep[][]{2004A&A...427....1E,2009ApJ...698..676D,2017MNRAS.468.2447P}} and to reveal high degree of polarisation of the observed emission at the NIR wavelengths {\citep[][]{2006A&A...455....1E,2015A&A...576A..20S}}. These findings favour synchrotron origin of the emission, in accordance with the predictions of the models associated with instabilities in an Advection-Dominated Accretion Flow (ADAF) as a triggering phenomenon behind them {\citep[e.g.][]{2010ApJ...725..450D}}.     

Episodes of much more dramatic brightening should have been experienced by Sgr~A* in the recent past, namely some $\sim100$ years ago, as is clearly evidenced by the X-ray ``echoes'' of this activity propagating across the Central Molecular Zone {\citep[CMZ,][for a review]{1993ApJ...407..606S,1996PASJ...48..249K,2013ASSP...34..331P}}. Key parameters (age, duration, spectral hardness and overall released energy) of the most recent flare(s) have been well constrained through sensitive observations of several molecular complexes, which happened to be bright in the reflected X-ray emission {\citep[][]{2004A&A...425L..49R,2007ApJ...656L..69M,2009PASJ...61S.241I,2010ApJ...714..732P,2012A&A...545A..35C,2013A&A...558A..32C,2013PASJ...65...33R,2015ApJ...815..132Z,2017MNRAS.468.2822K,2018A&A...610A..34C,2018A&A...612A.102T}}. Namely, we know that one of the recent flares occurred $\sim 120$ yrs ago, lasted for $\lesssim$1.5 yrs with the total emitted power $\sim10^{47}$ erg and hard X-ray spectrum {\citep[][]{2017MNRAS.465...45C}}.

The inferred shortness of this flare implies that only a very thin ($\lesssim0.5$ pc) shell of the dense gas confined by two parabolic surfaces is observed in X-ray reflection at any given moment\footnote{The parabolic shape is an approximation for the exact ellipsoidal one (with the observer and the primary source residing in its two foci) in the relative vicinity of the primary source \citep{1939AnAp....2..271C}.}, and one might readily simulate the expected evolution of the reflected signal over the light-crossing-time of the CMZ, given a certain model of dense gas distribution across it {\citep[][]{2017MNRAS.468..165C}}. In particular, one can make a prediction regarding the expected degree of polarisation in the reflected X-ray continuum, which is, in the simplest scenario, determined merely by the cosine of the scattering angle {\citep{2002MNRAS.330..817C,2017MNRAS.468..165C,2014MNRAS.441.3170M,2015A&A...576A..19M}}. Observations with the forthcoming imaging X-ray polarimeters \citep[e.g. \textit{IXPE,}][]{2016SPIE.9905E..17W} will allow checking these predictions and test the whole paradigm in an independent {and complementary} way {\citep[][]{2017MNRAS.471.3293C,2019arXiv191010092S}}. 

Here we consider an extension of the basic paradigm allowing for possible intrinsic polarisation of the original flare's emission. \citet{2002MNRAS.330..817C} considered illumination of a single cloud by linearly-polarised primary X-ray emission and presented explicit expressions for the polarisation degree of the scattered emission as a function of the cloud's relative position {\citep[see also][where scattering of anisotropic  and polarised time-variable radiation has been modelled]{1987SvAL...13..233G}}. We extend that analysis by considering the full polarisation pattern of the scattered emission, i.e. the maps of the polarisation degree and polarisation plane orientation for various configurations of the primary source polarisation. We show how the presence of the linearly-polarised component in the primary emission modifies morphological properties of the observed polarisation field and discuss diagnostic possibilities arising from this. These diagnostics could be already applied {to} the data of the upcoming X-ray polarimeter missions.

\section{Polarisation of the X-ray echo}
\label{s:reflection}
\begin{figure}
\centering
\includegraphics[width=1.0\columnwidth,viewport=60 340 530 700]{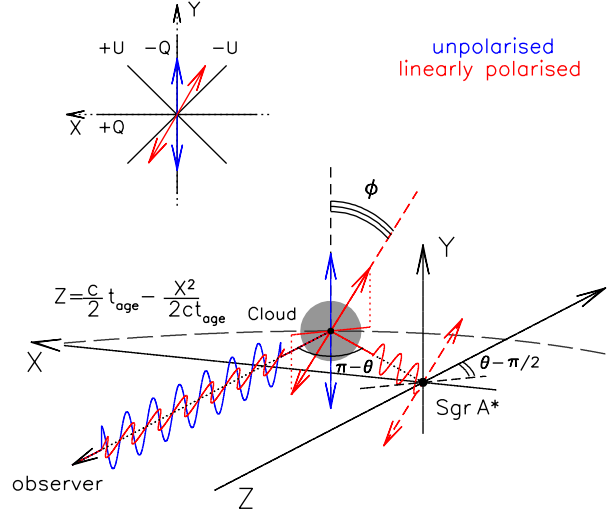}
\caption{A sketch illustrating polarisation of the X-ray emission arising due to scattering of an intrinsically unpolarised (blue) and linearly-polarised {(red)} radiation. {The coordinate system is chosen so that the primary source, Sgr~A*, is located in the origin, $z$-axis points from the observer towards it, and the $xz$-plane contains center of the reflecting cloud (grey sphere)}. The line-of-sight and picture plane positions of the cloud are connected by the parabolic relation { $Z=\frac{c}{2}t_{\rm age}-\frac{X^2}{2ct_{\rm age}}$}, where $t_{\rm age}$ is the age of the flare. For the intrinsically unpolarised primary source, the electric field vector of the reflected emission {(the solid blue arrow)} is perpendicular to the scattering (i.e. $xz$) plane and polarisation degree is set by the cosine of the scattering angle $\theta$. For the linearly-polarised primary source {(with the electric field vector depicted as the red arrows)}, the intensity, polarisation angle and polarisation degree all depend on the angle $\phi$ between the electric field vector of the primary emission and the {normal to the} scattering plane. The inset shows definition of the Stokes parameters used throughout the paper and predicted polarisation plane orientations for the two illustrated cases. 
}
\label{fig:sketch_pol}
\end{figure}

Let us consider a {small} cloud of the dense gas that is reflecting X-ray emission from Sgr~A*'s flare that occurred $t_{\rm age}\sim100$ yrs ago. Except for the photoionisation followed by fluorescent line emission (most importantly the fluorescent line of the neutral iron at 6.4 keV), the key process is Compton scattering, which, in majority of the relevant cases, might be treated in the optically thin limit \citep[e.g.][]{2017MNRAS.471.3293C}. The polarisation of the reflected X-ray continuum is fully determined by the geometry of the scattering, set by the relative disposition of the cloud and the primary source.

We use the coordinate system such that the primary source is located in the origin, the $z$-axis is pointing from the observer towards it, and the reflecting cloud resides in the $xz$-plane (see Fig.\ref{fig:sketch_pol}). This plane is referred to as the scattering plane in what follows, and the $y$-axis represents then the direction normal to it.

The line-of-sight and picture plane coordinates of the reflecting gas are connected by the relation 
\begin{equation}
    Z=\frac{c}{2}t_{\rm age}-\frac{X^2}{2ct_{\rm age}},
\end{equation}
which can be formulated in the dimensionless (and time-independent) form after dividing by the characteristic length scale $R_0=ct_{\rm age}\approx31$pc $(t_{\rm age}/100\,{\rm yr})$, so that $(x,y,z)=(X,Y,Z)/ct_{\rm age}$:
\begin{equation}
    {z}=\frac{1}{2}\left(1-x^2\right).
    \label{eq:paraboloid}
\end{equation}
The distance from the primary source to the cloud equals
\begin{equation}
    R/R_0=\sqrt{x^2+z^2}=\frac{1}{2}\left(1+x^2\right),
\end{equation}
and cosine of the scattering angle $\theta$ (see Fig.\ref{fig:sketch_pol}) is calculated as
\begin{equation}
\mu=\cos\theta=-Z/R=-\left(1-{x}^2\right)/\left(1+{x}^2\right).  
\end{equation}

For the flare duration $\Delta t\sim 1$ yr $\ll t_{\rm age}$, the thickness of the gas layer contributing to the reflected emission for any line-of-sight direction $(X,Y)$ is small 
\begin{equation}
    \Delta \mu\approx\frac{d\mu}{dt}\Delta t\sim \Delta t/t_{\rm age}\ll 1,
\end{equation}
so one can neglect variations in the scattering geometry across this layer.

Following \citet{2002MNRAS.330..817C}, let us now assume that the primary flare emission consists of an unpolarised component with intensity $a_0$ and a linearly-polarised one with intensity $a$.  If the electric vector of the polarised component makes an angle $\phi$ with the $y$-axis, the amplitudes of the electric field in the picture plane are $E_y\propto \cos \phi$ and $E_x\propto \mu\sin \phi$ (see Fig.\ref{fig:sketch_pol}).

The Stokes vector ${\bf S_p}$ \citep[see e.g.][and inset in Fig. \ref{fig:sketch_pol} for the definition]{2014JKAS...47...15T} of the scattered light corresponding to the linearly polarised component is 
\begin{align}
{\bf S_p}\equiv\begin{pmatrix} I \\ Q \\ U \end{pmatrix}=\begin{pmatrix} E^2_x+E^2_y \\ E^2_x-E^2_y \\ 2 E_xE_y \end{pmatrix}=a\begin{pmatrix} \mu^2\sin^2\phi   + \cos^2\phi \\ \mu^2\sin^2\phi-\cos^2\phi \\ 2\mu\sin\phi\cos\phi\end{pmatrix},
\end{align}
where we dropped the circular polarisation component $V$.

Accordingly, for the unpolarised component one has
\begin{align}
{\bf S_0}=\frac{a_0}{2}\begin{pmatrix} \mu^2+1 \\  \mu^2-1 \\ 0 \end{pmatrix},
\end{align}
as can be obtained by averaging out the previous expression over all possible electric field vector orientations.

The resulting total Stokes vector of the mixture of the two scattered components equals
\begin{equation}
{\bf S}={\bf S_p}+{\bf S_0}=
\begin{pmatrix} {a_0}\,(\mu^2+1)/2+a\,(\mu^2\sin^2\phi+\cos^2\phi)\\  {a_0}\,(\mu^2-1)/2+a\,(\mu^2\sin^2\phi-\cos^2\phi) \\2\,a\,\mu\sin\phi\cos\phi  \end{pmatrix}.
\label{eq:stokesscat}
\end{equation}

The observed degree of polarisation equals
\begin{align}
P_{\rm s}(\mu,\phi)= \frac{\sqrt{Q^2+U^2}}{I}=~~~~~~~~~~~~~~~~~~ \nonumber \\
~~~~~~~~~~~\frac{\sqrt{\left[1-\mu^2+P_0(1+\mu^2)\cos2\phi\right]^2+4P_0^2\mu^2\sin^2 2\phi}}{1+\mu^2+P_0(1-\mu^2)\cos 2\phi},
\label{eq:pol}    
\end{align}  
where $P_0=a/(a+a_0)$ is the polarisation degree of the primary radiation \citep[cf. eq.3 in ][]{2002MNRAS.330..817C}.

The polarisation plane is inclined by an angle
\begin{align}
\psi=\frac{1}{2}\arctan_2\left( \frac{U}{Q}\right )=-\frac{1}{2} \arctan_2 \left [ \frac{2 P_0 \mu \sin 2\phi}{(\mu^2-1)-P_0(\mu^2+1)\cos 2\phi} \right ]
\end{align}
relative to the $y$-axis, where  $\arctan_2$ is the quadrant-preserving arc tangent function.

{Combining this 'single-cloud' formalism with the scattering geometry set by the paraboloid relation Eq.~\ref{eq:paraboloid}}, one can readily calculate the full picture plane pattern of the polarisation for the scattered emission. {Namely, the transformation from the local single-cloud based coordinate system to some arbitrary global coordinate system centred on the primary source is achieved by simple rotations. The corresponding polarisation patterns are most easily obtained by the standard rotation transformations of the Stokes vectors given above.} Explicit Cartesian formulae for calculating the such a pattern for linearly-polarised primary emission with arbitrary orientation of the electric field vector are presented in Appendix \ref{s:a1}.

\begin{figure}
\centering
\includegraphics[width=0.83\columnwidth,viewport=40 190 550 670]{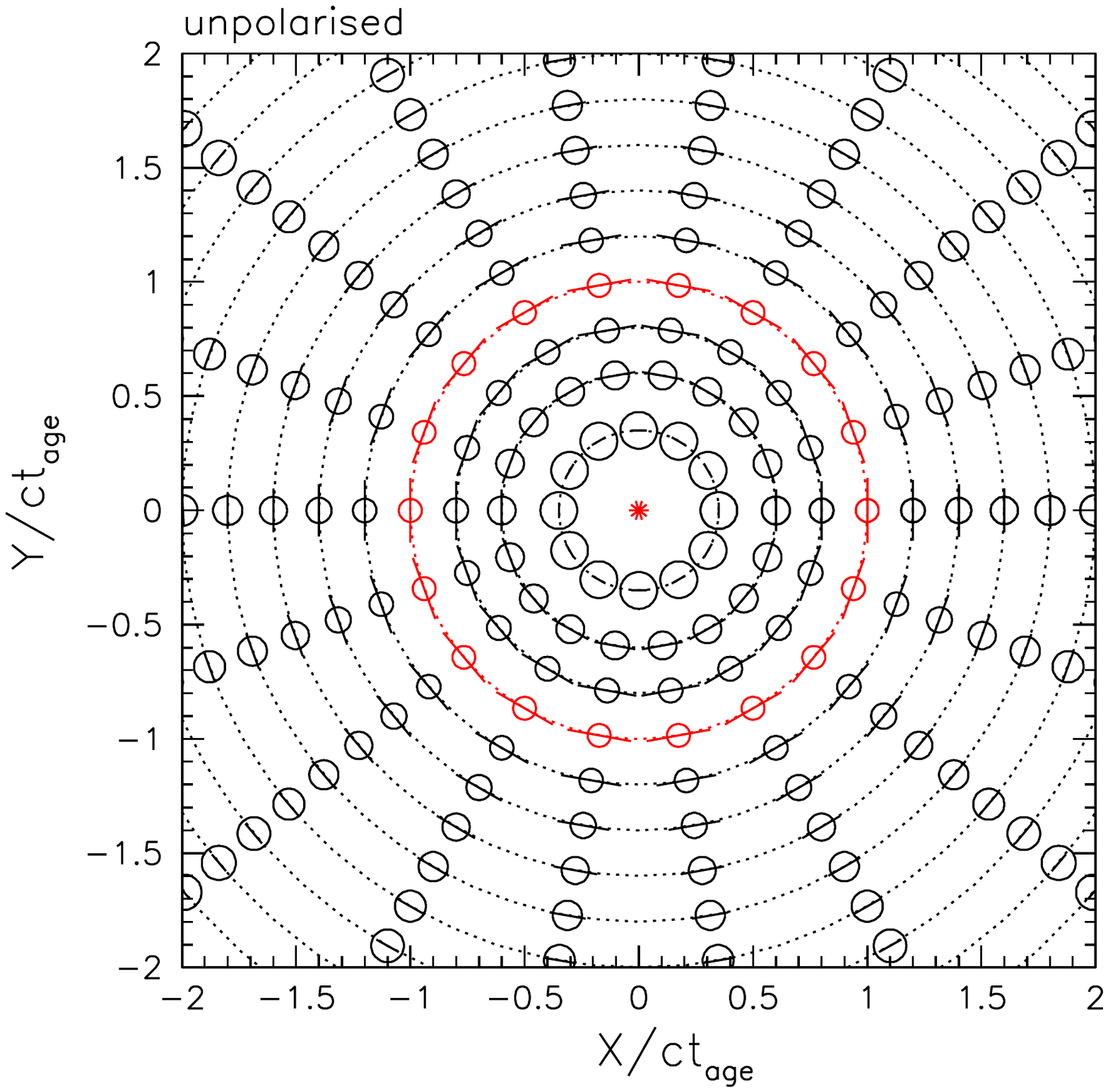}
\includegraphics[width=0.83\columnwidth,viewport=40 190 550 685]{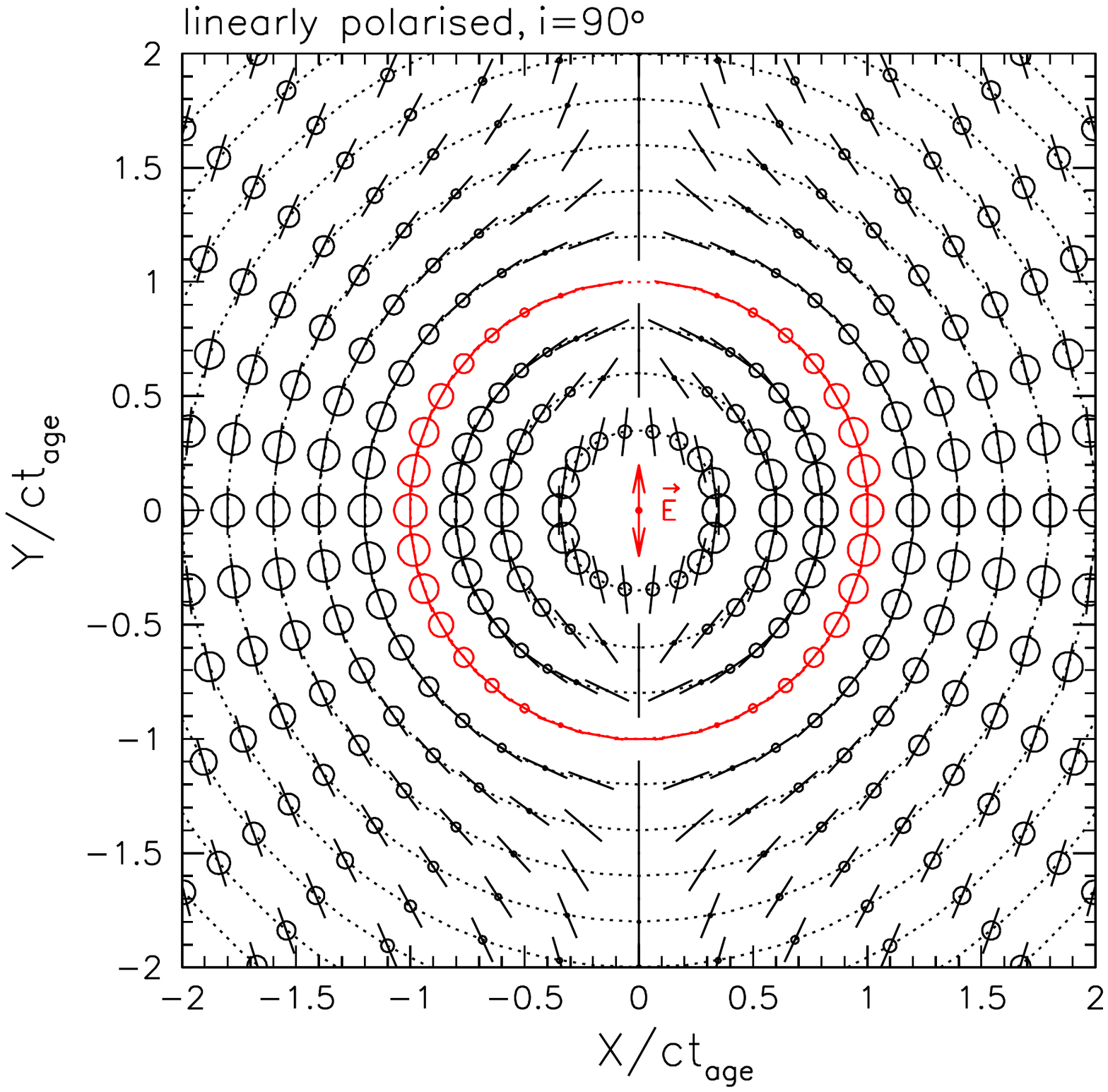}
\includegraphics[width=0.83\columnwidth,viewport=40 190 550 685]{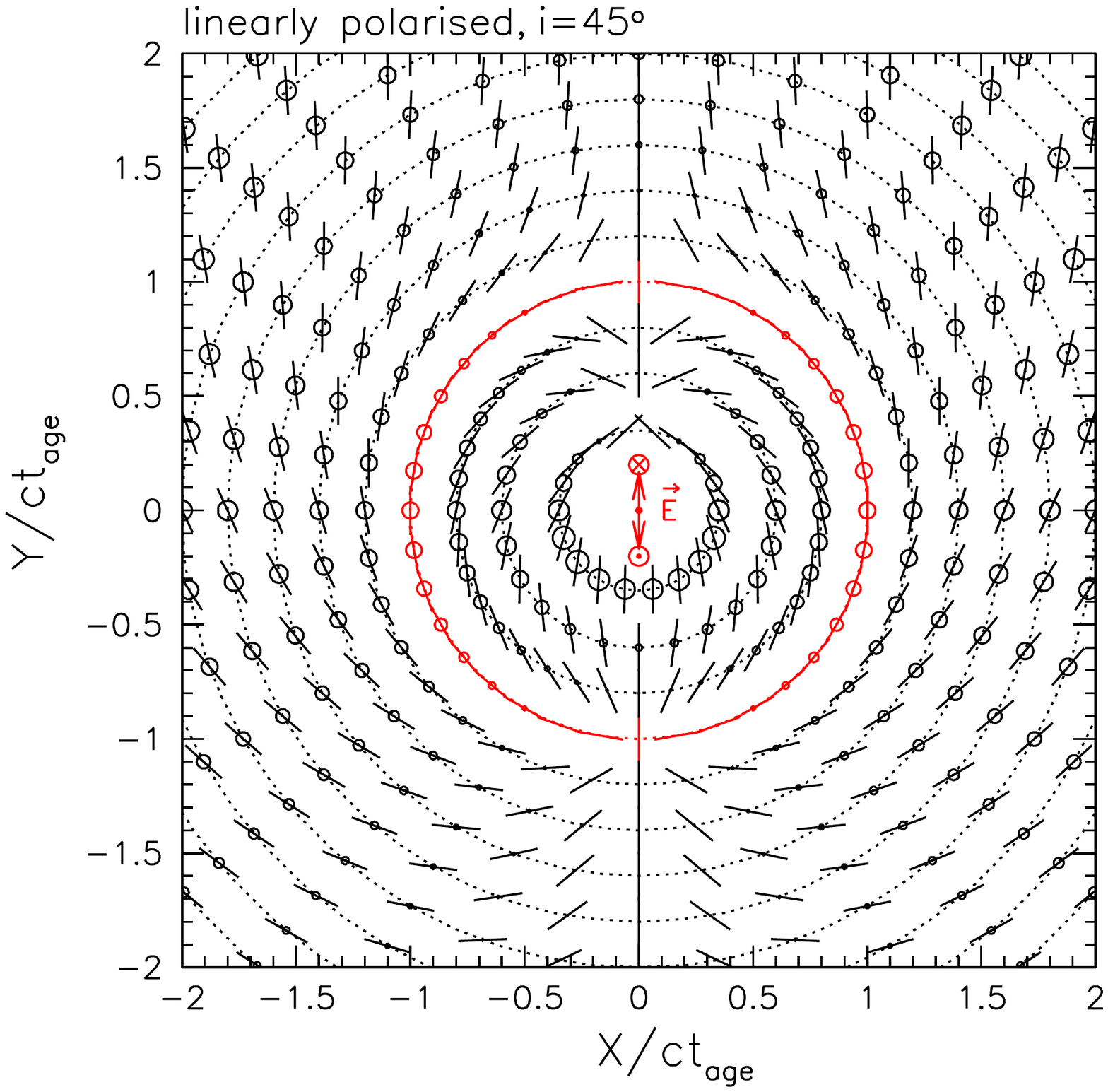}
\caption{Global {(defined relative to the primary source position rather than the scattering cloud)} polarisation patterns of the reflected X-ray continuum for the case of unpolarised (top panel) or linearly polarised (middle and bottom panels) primary emission from Sgr~A*. The coordinates are in dimensionless (and time-independent) form, with $R_0=ct_{\rm age}$ corresponding to unity (points on the $R=R_0$ circle are shown in red). Size of the circles scales with the relative intensity of the scattered emission, while the bars show polarisation plane orientation and polarisation degree (proportional to the bar's length). Projection of the electric field vector (inclined by angle $i$ with respect to the line-of-sight,{ the top side directed away from the observer}) on the picture plane is depicted in the centres of the plots. 
}
\label{fig:unpollinpol}
\end{figure}

In the case of unpolarised primary source, $P_0=0$, one has $\displaystyle P_{\rm s}(\mu,\phi)=\frac{1-\mu^2}{1+\mu^2}$ and $\psi=0$, i.e. the polarisation plane orientation is always perpendicular to the line connecting the source and the cloud (as shown in the top panel of Fig. \ref{fig:unpollinpol}). The phase function induces the intensity dependence of the scattered continuum in the form of $I_{\rm s}\propto (1+\mu^2)$. 

For the linearly-polarised primary emission, $P_0=1$, and, as can be most easily seen from Eq.\ref{eq:stokesscat}, $\displaystyle P_{\rm s}(\mu,\phi)=1$ everywhere as well. The pattern of the intensity in this case takes more complicated shape,  $I_{\rm s}\propto (\mu^2\,\sin^2\phi+\cos^2\phi)$, due to intrinsic anisotropy of the linearly-polarised emission. This effect is most significant for the positions from which the primary source is viewed nearly along the direction of the electric field vector $\vec{E}$.

A simple situation when the electric field vector $\vec{E}$\, is in the picture plane (i.e. inclined by $i=90\,\deg$ relative to the line-of-sight) is illustrated in the middle panel of Fig. \ref{fig:unpollinpol}. Due to the presence of the specific direction (set by the projection of $\vec{E}$) in this case, the polarisation pattern loses axial symmetry and acquires quadrupole-like shape (i.e. up-down vs. left-right asymmetry in this example).  For the directions nearly perpendicular to the electric field vector, the polarisation pattern (including scattering intensity) stays very similar to the case of unpolarised primary source. However, it changes dramatically for the directions close to the $\vec{E}$ axis. This change involves both scattered intensity (as mentioned above) and polarisation plane orientation, with the latter tending to align with the $\vec{E}$ projection.

Taking into account combination of these two effects, it is clear that perhaps the most promising directions for revealing presence of the linearly-polarised component in the primary emission would be those which are nearly aligned both with the primary source's direction and the projection of the $\vec{E}$ axis (i.e. points inside the $R=R_0=ct_{\rm age}$ (red circle) and close to the $y$-axis in Fig. \ref{fig:unpollinpol}). 

Somewhat more complicated pattern arises when the electric field vector is inclined {by an arbitrary angle} relative to the line-of-sight. Such a situation, with $\vec{E}$ inclined by $i=45\,\deg$, is illustrated in the bottom panel of Fig. \ref{fig:unpollinpol}. In this case, the pattern bears not only quadrupole-like asymmetry, but also a dipole one due to  non-orthogonality of the paraboloid axis and the symmetry axis of the primary emission. Here, as before, directions, which are nearly perpendicular to the projection of $\vec{E}$ on the picture plane, are not strongly affected. However, for the directions aligned with the projection of the $\vec{E}$ axis, the overall pattern is characterised by remarkable up-down asymmetry, both in the scattered intensity and polarisation plane orientation. Once again, the points inside $R=R_0=ct_{\rm age}$ circle (shown in red) offer the best opportunities to reveal such a pattern from the observational point of view. 

\begin{figure}
\centering
\includegraphics[width=0.95\columnwidth,viewport=40 200 580 680]{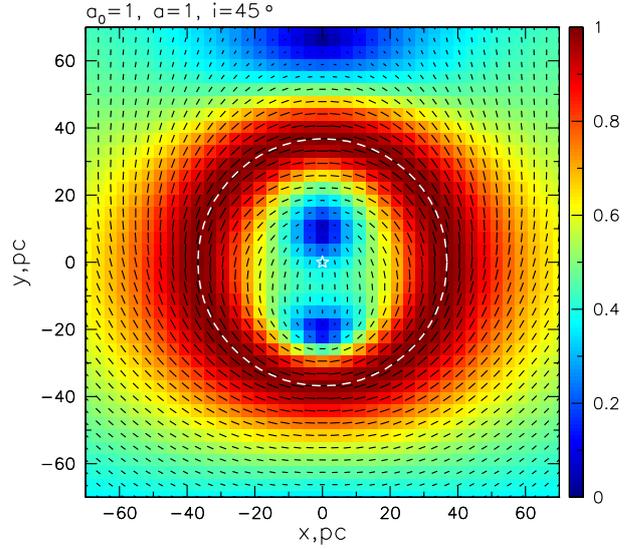}
\caption{Predicted polarisation degree (colour-coded) and polarisation plane orientation (black bars) of the reflected continuum in the case of an even mixture of unpolarised and linearly-polarised (with the electric field vector $\vec{E}$ inclined by $45\deg$ relative to the line-of-sight {with the picture plane projection aligned with the $y$-axis}) primary flare emission. { The primary source is located at the coordinate system origin, the age of the flare $t_{\rm age}$ is taken equal to $120$ yrs, which corresponds to the picture plane cross-section of the reflecting paraboloid (shown by the white circle) $R_0=ct_{\rm age}=37$ pc.}}
\label{fig:linpol_vert45}
\end{figure}

Finally, we present {an example when} the polarisation pattern of the scattered emission is an even mixture of the unpolarised and linearly- polarised (with the electric field vector $\vec{E}$ inclined by $i=45\deg$ relative to the line-of-sight) components. Such a situation is shown in Fig. \ref{fig:linpol_vert45}, where the coordinate axes now correspond to the physical scales assuming $t_{\rm age}=120$ yrs (i.e. $R_0=ct_{\rm age}=
37$ pc, shown by the white circle). Colour-coded is the resulting polarisation degree, while the black bars demonstrate orientation of the polarisation plane. In contrast to the simple axis-symmetric picture expected for the unpolarised primary source, this picture shows high level of complexity both in terms of the polarisation degree variations and morphology of the polarisation plane orientations.

\section{Discussion}
\label{s:discussion}

In the previous section we presented calculations that illustrate {global} polarisation patterns that arise when the emission of the X-ray flare contains a linearly-polarised component. The resulting patterns demonstrate substantial morphological complexity, in contrast to the case of the unpolarised primary emission.

Besides the change in the polarisation degree of the scattered emission, given by Eq.\ref{eq:pol} and already discussed by \citealt{2002MNRAS.330..817C}, there is a very visual change in the orientation of the polarisation plane. This change is naturally connected with the preferred direction set by the projection of the electric field vector of the linearly-polarised primary emission on the picture plane (or, more accurately, on the surface of the reflecting paraboloid).

Presence of the linearly-polarised component in the primary emission modifies a simple prediction that the polarisation plane is always perpendicular to the line connecting the source and the reflecting cloud. Hence, one should be aware of this possibility when trying to reconstruct location of the primary source based on the polarisation plane orientations.{ Conversely, if the location of the primary source is confirmed, the data can be used to infer the intrinsic polarisation of the flare emission.} The clouds located on the {farther} side of the paraboloid (or, equivalently, inside the $R=ct_{\rm age}$ circle) are best suited for probing the intrinsic polarisation of the flare's emission, since only very low polarisation is expected for them in the case of unpolarised primary source. Also, the polarisation plane orientation for these directions is very close to the projection of the electric field vector on the picture plane.

Another important impact of the polarised primary emission is the induced modulation of the scattered intensity. Contrary to the polarisation plane orientation, however, revealing such a feature should be much more complicated due to inhomogeneity of the scattering medium, which might have a certain systematic pattern as well (for instance following the Galactic plane). Still, certain symmetries inherent to the polarisation-induced modulation, in combination with the information on the polarisation plane orientation and the polarisation degree, could allow disentangling these effects, given sufficiently broad coverage of the directions in picture plane.

Intrinsic anisotropy of the primary emission, e.g. in the form of beaming, can also produce global modulation pattern of the scattered emission. In this case, however, one can use the equivalent width of the fluorescent lines as a discriminating probe. Indeed, intensity of the fluorescent line is virtually insensitive to the polarisation of the primary emission and geometry of the scattering \citep{1998MNRAS.297.1279S,2002MNRAS.330..817C}. Hence, combining the polarimetric measurements with the measurement of the iron fluorescent line's equivalent width could allow distinguishing intensity modulation caused by beaming of unpolarised source from the effect of linear polarisation of the primary emission. Substantial global variations of the fluorescent line's equivalent width might be considered as a possible indication of the polarised primary emission. Of course, there are other effects to be considered here, including variations in abundance of heavy elements, presence of unrelated background emission and possible contribution of the second scatterings \citep[][]{1998MNRAS.297.1279S,2002MNRAS.330..817C,2016A&A...589A..88M,2017MNRAS.468..165C,2019arXiv190611579K}. In principle, one might take advantage of the temporal behaviour expected for the reflected X-ray emission - it should not change dramatically over the light-crossing time of the cloud, but almost fully disappear after the cloud is fully 'scanned'. {The resulting polarisation pattern might take even more complicated shape determined by asymmetries in the gas density distribution and anisotropy of illumination \citep[see e.g.][where impact of multiple scatterings has been considered in the simple narrow beam approximation for the incident radiation field]{2002MNRAS.330..817C}.  
Although this situation might be indeed relevant for a very massive molecular complex like Sgr B2, we defer an in-depth discussion of the concrete observational methodology for such studies for future work \citep[see also][ for a relevant discussion]{2016A&A...589A..88M}.}


{We note in passing that some of the polarisation patterns considered above are reminiscent of the E-B-mode components widely used in application to the Cosmic Microwave Background \citep[e.g.][]{1997PhRvD..55.7368K,1997PhRvD..55.1830Z}. However, given the complicated mixture of factors affecting the spectra and polarisation properties of the reflected emission from the GC region, it is not obvious if the E-B decomposition could be efficiently applied here as well \citep[see][for a relevant discussion]{2001PhRvD..64j3001Z}.}

\begin{figure}
\centering
\includegraphics[width=0.9\columnwidth,viewport=20 160 570 700]{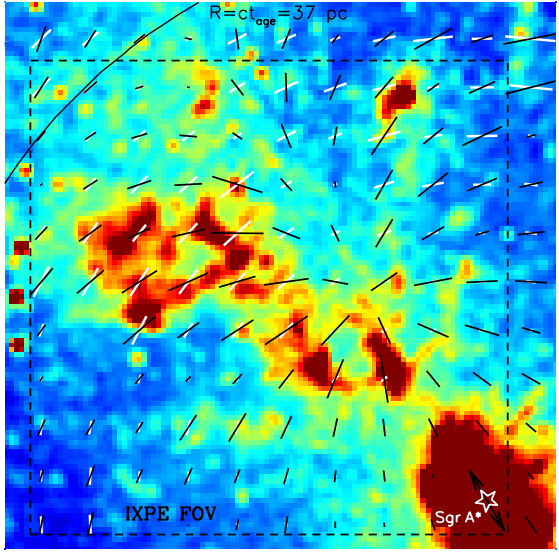}
\caption{An illustration of the polarisation pattern of the reflected continuum in the case of the linearly-polarised primary source (black bars) and the unpolarised one (white bars), based on the map of the reflected X-ray emission in the region of the Sgr~A molecular complex provided by $Chandra$. Field-of-view of the forthcoming imaging X-ray polarimeter $IXPE$ is depicted as the black 12.9'x12.9' square. The position of Sgr~A* and projection of the electric field vector (arbitrarily selected with inclination 45 deg relative to the line-of-sight) are shown. The age of the flare $t_{\rm age}=120$ yrs, the black solid line is part of the $R=ct_{\rm age}\approx37$ pc circle.}
\label{fig:chandraixpe}
\end{figure}

An important limitation of the X-ray reflection data arises from the fact that only the currently brightest (in scattered emission) molecular complexes might be accessible for the imaging polarimetry due to relatively low sensitivity and high level of contaminating background in the Galactic Centre region. As a result, only small patches of the patterns considered above will have a chance to be observed {at a given moment}. In order to illustrate prospects of the forthcoming imaging polarimeter mission $IXPE$ \citep{2016SPIE.9905E..17W}, we combine an example of the predicted polarisation pattern with the map of the reflected X-ray emission inferred for the Sgr~A molecular complex from the data of $Chandra$ observatory \citep[see][for details]{2017MNRAS.465...45C}. The reflected X-ray emission from this complex is currently the brightest one across the CMZ\footnote{The giant molecular complex Sgr~B2 was considered as the most promising target thanks to its mass, size and plausibly favourable disposition { \citep{2002MNRAS.330..817C,2014MNRAS.441.3170M,2015A&A...576A..19M}}. However, the reflected X-ray emission from it has faded away as the echo had fully scanned it.} and, under the standard assumptions, polarisation signal from it is expected to be detected with high ($\gtrsim10\sigma$) significance after a 1Ms exposure with $IXPE$. In Fig. \ref{fig:chandraixpe} we show that the sizes of $IXPE$'s field-of-view (12.9 arcmin square) and the molecular complex itself could in principle allow distinguishing the polarisation pattern arising from the linearly-polarised primary from source (black bars) and from the unpolarised one (white bars){, albeit the assumed 100\% linear polarisation and favourable orientation of its electric field vector are clearly an overly extreme.} 

Finally, we comment on the implications that measurements of polarisation of the primary flare emission might have on revealing their physical origin. Clearly, it will be crucial to compare the inferred polarisation properties with those measured (with ever-improving quality) for the much more frequent and orders of magnitude fainter NIR and X-ray flares \citep[e.g.][for a review]{2010RvMP...82.3121G}. It has been proposed that the historical flares might represent an extreme end of the same population, i.e. they might share the same underlying physical mechanism \citep{2012ApJS..203...18W,2013ApJ...774...42N,2015MNRAS.453..172P,2017MNRAS.468..165C}. In that case, comparing polarisation properties will allow checking predictions for the models invoking stochastic changes in the (low) mass accretion rate \citep[e.g.][]{2000ApJ...545..842Q}. On the other hand, the historical flares might be caused by completely unrelated events, leading to episodes of radiatively efficient accretion \citep[e.g.][]{2008MNRAS.383..458C,2012MNRAS.421.1315Z,2019MNRAS.486.1833S}. In such situation, the polarisation of the primary emission also might arise, resulting from relativistic effects and Comptonisation of the disk emission \citep{1980ApJ...235..224C,1985A&A...143..374S}. Finally, the flare emission might be connected to (beamed) radiation of a relativistic jet \citep[e.g.][]{2000A&A...362..113F}. All these models have their distinct signatures \citep[which could also be intimately connected with the spin of Sgr~A*, e.g.][]{2008MNRAS.391...32D}, and some of them might be tested against the forthcoming polarimetric data. Various models predict a certain level of both anisotropy and polarisation of the primary emission. The X-ray echoes could in principle allow reconstructing the full 3D dependence of the primary emission's intensity and polarisation. The simple framework presented here offers a minimal extension of the basic model, which can be exploited for constructing more sophisticated physically-motivated models {\citep[see e.g.][for a consideration of scattering of polarised and anisotropic radiation in application to time-variable radio sources in clusters of galaxies]{1987SvAL...13..233G}}. 

{Throughout the paper we assumed that Sgr~A* was indeed the source of the primary emission. The possibility that the flare originated from some other (transient) source in the Galactic Centre region cannot be {entirely} excluded yet. All properties of the reflected emission, including polarisation, are virtually identical if the primary source is located within $\sim$ a few pc of Sgr~A* \citep[as e.g. is the magnetar SGR~J1745-29, see][]{2013ApJ...775L..34R}, making this case indistinguishable from the basic one {(provided that rather severe constraints on the luminosity and fluence are satisfied)}.  For other locations, the expected polarisation pattern can be obtained from the presented patterns by displacement of their centre from Sgr~A* to the actual position of the source and adjustment of the flare's parameters (primarily, the age). For the case of unpolarised primary emission, the accuracy of the primary source's position determination from the polarisation plane orientation has been discussed before  \citep[in application to the forthcoming X-ray polarimeters][]{2015A&A...576A..19M,2017MNRAS.468..165C}. The complex picture arising in the case of the polarised primary source implies that a more intricate analysis involving spectroscopic and polarisation measurements for several molecular complexes will be required. Using the inherent symmetries of the predicted patterns (akin E-B mode decomposition) might turn out to be useful in this case.} 

\section{Conclusions}
\label{s:conclusions}

{One of the major objectives of the forthcoming imaging polarimeter missions will be the detection of the polarised emission induced by scattering of the putative X-ray flare(s) from Sgr~A* on molecular clouds. Once such polarised emission is found, the role of Sgr~A* as a primary source can be confirmed and the estimates of the age of the flare will be improved (or the presence of several flares separated by tens or hundreds of years will be revealed). The above consideration is largely based on the assumption that the primary emission is unpolarised. In this study, we have developed a simple framework for calculation of the intensity/polarisation spatial patterns for the case when the flare emission is polarised itself.

We show that the resulting patterns demonstrate substantial morphological complexity, in contrast to the case of the unpolarised primary emission. It includes non-axisymmetric variations in the intensity, polarisation degree and polarisation plane orientation, which in this case is not necessarily perpendicular to the line connecting the primary source and the reflecting cloud. Positions corresponding to the further side of the reflecting paraboloid, i.e. projected inside the characteristic radius $R=ct_{\rm age}$ offer the best opportunities to reveal intrinsic polarisation of the flare's emission.

We notice the possibility of combining the polarimetric data with the spectroscopic measurements of equivalent width of the fluorescent lines in order to probe possible anisotropy of the flare's emission. Application of the outlined approaches to the real data might become feasible already with the forthcoming imaging X-ray polarimeters (e.g. $IXPE$) and is only limited by the presence/absence of dense enough clouds around Sgr~A*.} The inferred polarisation (and possibly beaming) properties of the historical flares will help in clarifying their physical origin and putting them in the general context of Sgr~A* accretion flow phenomenology.

\section*{Data availability}
No new data were generated or analysed in support of this research.

\section*{Acknowledgements}

{IK, EC, RS} acknowledge support by the Russian Science Foundation grant~19-12-00369. We are grateful to the anonymous referee for the useful suggestions that allowed to improve the paper.


\newpage
\appendix
\section{Explicit Cartesian expressions for scattered linearly-polarised emission}
\label{s:a1}

Let us assume that the direction of the initial polarisation vector is given by $\vec{E}=(e_x,e_y,e_z)$ with respect to some fixed coordinate system set by the line-of-sight direction and a coordinate system in the picture plane (similar to the one illustrated in Figure \ref{fig:sketch_pol}, but not necessarily related to any particular reflecting cloud). Consider now a point with coordinates $(X,Y,Z)$ in the same coordinate system which contributes to the scattered flare emission at the given moment, i.e. lying on the paraboloid { $Z=\frac{ct_{\rm age}}{2}-\frac{X^2+Y^2}{2ct_{\rm age}}$.}  
After reducing the coordinates to $(x,y,z)=(X,Y,Z)/ct_{\rm age}$, the whole picture does not depend on time anymore.

In these coordinates, the electric field vector of the scattered emission to be observed is given by the $x-$ and $y-$components
\begin{align}
s_x(x,y)=\frac{1}{\left(1+x^2+y^2\right)^2}~\times& 
\left[\,\left(1-x^2+y^2\right)\left(1+3x^2+y^2\right)\,e_x\right.     \nonumber\\
&-4xy\,e_y\nonumber\\
&+\left.\,2x(x^2+y^2-1)\,e_z\,\right]  \label{eq:a1}\\
s_y(x,y)=\frac{1}{\left(1+x^2+y^2\right)^2}~\times&
\left[\,\,-4xy\,e_x\right.\nonumber\\
&+\left(1+x^2-y^2\right)\left(1+x^2+3y^2\right)\,e_y\nonumber\\
&+\left.2y(x^2+y^2-1)\,e_z\,\right]. 
\label{eq:a2}
\end{align}

The corresponding Stokes parameters (defined as shown in the inset of Fig.\ref{fig:sketch_pol}) are calculated as usual
\begin{align}
{\bf S_l}(x,y)\equiv\begin{pmatrix} I_{l}(x,y) \\ Q_l(x,y) \\ U_l(x,y) \end{pmatrix}=\begin{pmatrix} s^2_x+s^2_y \\ s^2_x-s^2_y \\ 2 s_x\,s_y \end{pmatrix}.
\label{eq:a3}
\end{align}

The set of Eqs. \ref{eq:a1}, \ref{eq:a2} and \ref{eq:a3} presents the full solution of the problem in explicit Cartesian form.

\bsp	
\label{lastpage}

\begin{thebibliography}{99}
\bibitem[\protect\citeauthoryear{Baganoff, et al.}{2001}]{2001Natur.413...45B} Baganoff F.~K., et al., 2001, Nature, 413, 45

\bibitem[\protect\citeauthoryear{Boyce, et al.}{2019}]{2019ApJ...871..161B} Boyce H., et al., 2019, ApJ, 871, 161
%
\bibitem[\protect\citeauthoryear{Capelli, et al.}{2012}]{2012A&A...545A..35C} Capelli R., Warwick R.~S., Porquet D., Gillessen S., Predehl P., 2012, A\&A, 545, A35
%
\bibitem[\protect\citeauthoryear{Chuard, et al.}{2018}]{2018A&A...610A..34C} Chuard D., et al., 2018, A\&A, 610, A34
%
\bibitem[Churazov et al.(2002)]{2002MNRAS.330..817C} Churazov, E., Sunyaev, R., \& Sazonov, S.\ 2002, \mnras, 330, 817
%
\bibitem[Churazov et al.(2017a)]{2017MNRAS.465...45C} Churazov, E., Khabibullin, I., Sunyaev, R., et al.\ 2017, \mnras, 465, 45
\bibitem[Churazov et al.(2017b)]{2017MNRAS.468..165C} Churazov, E., Khabibullin, I., Ponti, G., et al.\ 2017, \mnras, 468, 165
\bibitem[Churazov et al.(2017c)]{2017MNRAS.471.3293C} Churazov, E., Khabibullin, I., Sunyaev, R., et al.\ 2017, \mnras, 471, 3293
%
\bibitem[\protect\citeauthoryear{Clavel et al.}{2013}]{2013A&A...558A..32C} Clavel M., Terrier R., Goldwurm A., Morris M.~R., Ponti G., Soldi S., Trap G., 2013, A\&A, 558, A32 

\bibitem[\protect\citeauthoryear{Connors, Piran \& Stark}{1980}]{1980ApJ...235..224C} Connors P.~A., Piran T., Stark R.~F., 1980, ApJ, 235, 224

\bibitem[\protect\citeauthoryear{Couderc}{1939}]{1939AnAp....2..271C} Couderc P., 1939, Annales d'Astrophysique, 2, 271
%
\bibitem[\protect\citeauthoryear{Cuadra, Nayakshin \& Martins}{2008}]{2008MNRAS.383..458C} Cuadra J., Nayakshin S., Martins F., 2008, MNRAS, 383, 458
%
\bibitem[\protect\citeauthoryear{Dodds-Eden, et al.}{2009}]{2009ApJ...698..676D} Dodds-Eden K., et al., 2009, ApJ, 698, 676
%
\bibitem[\protect\citeauthoryear{Dodds-Eden, et al.}{2010}]{2010ApJ...725..450D} Dodds-Eden K., Sharma P., Quataert E., Genzel R., Gillessen S., Eisenhauer F., Porquet D., 2010, ApJ, 725, 450
%
\bibitem[\protect\citeauthoryear{Dov{\v{c}}iak, et al.}{2008}]{2008MNRAS.391...32D} Dov{\v{c}}iak M., Muleri F., Goosmann R.~W., Karas V., Matt G., 2008, MNRAS, 391, 32
%
\bibitem[\protect\citeauthoryear{Eckart, et al.}{2004}]{2004A&A...427....1E} Eckart A., et al., 2004, A\&A, 427, 1
\bibitem[\protect\citeauthoryear{Eckart, et al.}{2006}]{2006A&A...455....1E} Eckart A., Sch{\"o}del R., Meyer L., Trippe S., Ott T., Genzel R., 2006, A\&A, 455, 1

\bibitem[\protect\citeauthoryear{Falcke \& Markoff}{2000}]{2000A&A...362..113F} Falcke H., Markoff S., 2000, A\&A, 362, 113

\bibitem[\protect\citeauthoryear{Genzel, et al.}{2003}]{2003Natur.425..934G} Genzel R., et al., 2003, Natur, 425, 934

\bibitem[\protect\citeauthoryear{Genzel, Eisenhauer \& Gillessen}{2010}]{2010RvMP...82.3121G} Genzel R., Eisenhauer F., Gillessen S., 2010, RvMP, 82, 3121

\bibitem[\protect\citeauthoryear{Gilfanov, Sunyaev \& Churazov}{1987}]{1987SvAL...13..233G} Gilfanov M.~R., Sunyaev R.~A., Churazov E.~M., 1987, SvAL, 13, 233

\bibitem[\protect\citeauthoryear{Haggard, et al.}{2019}]{2019ApJ...886...96H} Haggard D., et al., 2019, ApJ, 886, 96

\bibitem[\protect\citeauthoryear{Inui, et al.}{2009}]{2009PASJ...61S.241I} Inui T., Koyama K., Matsumoto H., Tsuru T.~G., 2009, PASJ, 61, S241

\bibitem[\protect\citeauthoryear{Kamionkowski, Kosowsky \& Stebbins}{1997}]{1997PhRvD..55.7368K} Kamionkowski M., Kosowsky A., Stebbins A., 1997, PhRvD, 55, 7368

\bibitem[\protect\citeauthoryear{Khabibullin, et al.}{2019}]{2019arXiv190611579K} Khabibullin I., Churazov E., Sunyaev R., Federrath C., Seifried D., Walch S., 2019, arXiv, arXiv:1906.11579

\bibitem[Koyama et al.(1996)]{1996PASJ...48..249K} Koyama, K., Maeda, Y., Sonobe, T., et al.\ 1996, \pasj, 48, 249

\bibitem[\protect\citeauthoryear{Krivonos, et al.}{2017}]{2017MNRAS.468.2822K} Krivonos R., et al., 2017, MNRAS, 468, 2822

\bibitem[\protect\citeauthoryear{Marin, et al.}{2014}]{2014MNRAS.441.3170M} Marin F., Karas V., Kunneriath D., Muleri F., 2014, MNRAS, 441, 3170
\bibitem[\protect\citeauthoryear{Marin, et al.}{2015}]{2015A&A...576A..19M} Marin F., Muleri F., Soffitta P., Karas V., Kunneriath D., 2015, A\&A, 576, A19

\bibitem[\protect\citeauthoryear{Molaro, Khatri \& Sunyaev}{2016}]{2016A&A...589A..88M} Molaro M., Khatri R., Sunyaev R.~A., 2016, A\&A, 589, A88

\bibitem[\protect\citeauthoryear{Muno, et al.}{2007}]{2007ApJ...656L..69M} Muno M.~P., Baganoff F.~K., Brandt W.~N., Park S., Morris M.~R., 2007, ApJL, 656, L69

\bibitem[\protect\citeauthoryear{Neilsen, et al.}{2013}]{2013ApJ...774...42N} Neilsen J., et al., 2013, ApJ, 774, 42

\bibitem[\protect\citeauthoryear{Ponti et al.}{2010}]{2010ApJ...714..732P} Ponti G., Terrier R., Goldwurm A., Belanger G., Trap G.,2010,ApJ,714,732 

\bibitem[\protect\citeauthoryear{Ponti et al.}{2013}]{2013ASSP...34..331P} Ponti G., Morris M.~R., Terrier R., Goldwurm A., 2013, ASSP, 34, 331  

\bibitem[\protect\citeauthoryear{Ponti, et al.}{2015}]{2015MNRAS.453..172P} Ponti G., et al., 2015, MNRAS, 453, 172

\bibitem[\protect\citeauthoryear{Ponti, et al.}{2017}]{2017MNRAS.468.2447P} Ponti G., et al., 2017, MNRAS, 468, 2447

\bibitem[\protect\citeauthoryear{Rea, et al.}{2013}]{2013ApJ...775L..34R} Rea N., et al., 2013, ApJL, 775, L34

\bibitem[\protect\citeauthoryear{Revnivtsev et al.}{2004}]{2004A&A...425L..49R} Revnivtsev M.~G., et al., 2004, A\&A, 425, L49 

\bibitem[\protect\citeauthoryear{Ryu, et al.}{2013}]{2013PASJ...65...33R} Ryu S.~G., et al., 2013, PASJ, 65, 33

\bibitem[\protect\citeauthoryear{Quataert \& Gruzinov}{2000}]{2000ApJ...545..842Q} Quataert E., Gruzinov A., 2000, ApJ, 545, 842

\bibitem[\protect\citeauthoryear{Sacchi \& Lodato}{2019}]{2019MNRAS.486.1833S} Sacchi A., Lodato G., 2019, MNRAS, 486, 1833

\bibitem[\protect\citeauthoryear{Shahzamanian, et al.}{2015}]{2015A&A...576A..20S} Shahzamanian B., et al., 2015, A\&A, 576, A20

\bibitem[\protect\citeauthoryear{Soffitta, et al.}{2019}]{2019arXiv191010092S} Soffitta P., et al., 2019, ESA Voyage 2050 White Paper, arXiv:1910.10092

\bibitem[\protect\citeauthoryear{Sunyaev \& Titarchuk}{1985}]{1985A&A...143..374S} Sunyaev R.~A., Titarchuk L.~G., 1985, A\&A, 143, 374

\bibitem[Sunyaev et al.(1993)]{1993ApJ...407..606S} Sunyaev, R.~A., Markevitch, M., \& Pavlinsky, M.\ 1993, \apj, 407, 606

\bibitem[\protect\citeauthoryear{Sunyaev \& Churazov}{1998}]{1998MNRAS.297.1279S} Sunyaev R., Churazov E., 1998, MNRAS, 297, 1279

\bibitem[\protect\citeauthoryear{Terrier, et al.}{2018}]{2018A&A...612A.102T} Terrier R., et al., 2018, A\&A, 612, A102

\bibitem[\protect\citeauthoryear{Trippe}{2014}]{2014JKAS...47...15T} Trippe S., 2014, JKAS, 47, 15

\bibitem[\protect\citeauthoryear{Weisskopf, et al.}{2016}]{2016SPIE.9905E..17W} Weisskopf M.~C., et al., 2016, SPIE, 9905, 990517, SPIE.9905

\bibitem[\protect\citeauthoryear{Witzel, et al.}{2012}]{2012ApJS..203...18W} Witzel G., et al., 2012, ApJS, 203, 18

\bibitem[\protect\citeauthoryear{Zaldarriaga \& Seljak}{1997}]{1997PhRvD..55.1830Z} Zaldarriaga M., Seljak U., 1997, PhRvD, 55, 1830

\bibitem[\protect\citeauthoryear{Zaldarriaga}{2001}]{2001PhRvD..64j3001Z} Zaldarriaga M., 2001, PhRvD, 64, 103001

\bibitem[\protect\citeauthoryear{Zhang, et al.}{2015}]{2015ApJ...815..132Z} Zhang S., et al., 2015, ApJ, 815, 132

\bibitem[\protect\citeauthoryear{Zubovas, Nayakshin, \& Markoff}{2012}]{2012MNRAS.421.1315Z} Zubovas K., Nayakshin S., Markoff S., 2012, MNRAS, 421, 1315 
\end{thebibliography}
\end{document}